\documentclass[reprint,preprintnumbers,amsmath,amssymb,aps,pra]{revtex4-1}
\newcommand{\half}{\mbox{$\frac{1}{2}$}}
\usepackage{amsmath}
\usepackage{amssymb}
\begin{document}
\title{Complex modes in unstable quadratic bosonic forms}
\author{R. Rossignoli, A.M. Kowalski} \affiliation{Departamento de F\'{\i}sica,
Universidad Nacional de La Plata, C.C.67, La Plata (1900), Argentina}
\begin{abstract}
We discuss the necessity of using non-standard boson operators for
diagonalizing quadratic bosonic forms which are not positive definite and its
convenience for describing the temporal evolution of the system. Such operators
correspond to non-hermitian coordinates and momenta and are associated with
complex frequencies. As application, we examine a bosonic version of a BCS-like
pairing Hamiltonian, which, in contrast with the fermionic case, is stable just
for limited values of the gap parameter and requires the use of the present
extended treatment for a general diagonal representation. The dynamical
stability of such forms and the occurrence of non-diagonalizable cases are also
discussed.
\end{abstract}
 \maketitle

Quadratic bosonic forms arise naturally in many areas of physics at different
levels of approximation. Starting from the basic example of coupled harmonic
oscillators, their ubiquity is testified by their appearance in standard
treatments of quantum optics \cite{Sa.91}, disordered systems \cite{GC.02},
Bose-Einstein condensates \cite{GM.97,LPBE.97,PB.97,AK.02} and other
interacting many-body boson and fermion systems \cite{RS.80,BR.86}. In the
latter they constitute the core of the random-phase approximation (RPA), which
arises as a first order treatment in a bosonized description of the system
excitations, or alternatively, from the linearization of the time-dependent
mean field equations of motion (time dependent Hartree, Hartree-Fock (HF) or
HF-Bogoliubov (HFB) \cite{RS.80,BR.86}). The ensuing forms are quite general
and may contain all types of mixing terms ($q_ip_j$, $q_iq_j$ and $p_ip_j$)
when expressed in terms of coordinates and momenta. Although the standard
situation, i.e., that where the RPA is constructed upon a stable mean field
(the Hartree, HF or HFB vacuum), corresponds to a positive form, in more
general treatments the RPA can also be made on top of unstable mean fields, as
occurs in the study of instabilities in binary Bose-Einstein condensates
\cite{GM.97,LPBE.97,PB.97,AK.02}, and even around non-stationary running mean
fields, as in the case of the static path + RPA treatment of the partition
function \cite{SPA1,SPA2}, derived from its path integral representation. In
these cases the ensuing forms may not be positive and may lead, as is well
known, to {\it complex} frequencies. Quadratic bosonic forms are also relevant
in the study of dynamical systems \cite{Do.00,Ko.00,P.97}, providing a basic
framework for investigating diverse aspects such as integrals of motion and
semiclassical limits.

Now, a basic problem with such forms is that while in the fermionic case they
can always be diagonalized by means of a standard Bogoliubov transformation
\cite{RS.80}, in the bosonic case they may not admit a similar diagonal
representation in terms of standard boson operators, nor in terms of usual
hermitian coordinates and momenta. These cases can of course only arise in
unstable forms which are not positive definite. The aim of this work is to
discuss the diagonal representation of such forms in terms of non-standard
boson-like quasiparticle operators (or equivalently, non-hermitian coordinates
and momenta), associated with complex normal modes.  This requires the use of
generalized Bogoliubov transformations since the usual one leads to a vanishing
norm in the case of complex frequencies. The present treatment allows then to
identify the operators characterized by an exponentially increasing or
decreasing evolution, providing a precise description of the dynamics and of
the quadratic invariants in the presence of instabilities. It will also become
apparent that an analysis of the dynamical stability based just on the
Hamiltonian positivity may not be sufficient.

As application, we will discuss a bosonic version of a BCS-type pairing
Hamiltonian, which, in contrast with the fermionic case, exhibits a complex
behavior, loosing its positive definite character above a certain threshold
value of the gap parameter, and becoming dynamically unstable above a second
higher threshold. In the presence of a perturbation it may even lead to a
reentry of dynamical stability after an initial breakdown. This example
illustrates the existence of simple quadratic forms which cannot be written in
diagonal form in terms of standard boson operators or coordinates and momenta.
Moreover, it also shows the existence of non-diagonalizable cases which do not
correspond to a zero frequency (and hence to a free particle term, in contrast
with standard Goldstone or zero frequency RPA modes arising from mean fields
with broken symmetries \cite{RS.80}), and which are characterized by equations
of motions which cannot be fully decoupled.

A general hermitian quadratic form in boson annihilation and creation operators
$b_i$, $b^\dagger_i$, can be written as
\begin{subequations}\label{H}
 \begin{eqnarray}
\!\!\!\!H&=&\sum_{i,j}A_{ij}(b^\dagger_i b_j+\half\delta_{ij}) + \half (B_{ij}
b^\dagger_i b^\dagger_j + B^*_{ij} b_i b_j) \label{Hbos}\\
&=&\half Z^\dagger \, \mathcal{H} \,  Z\,,   \hskip .2cm
 \mathcal{H}=\left(\begin{array}{cc}A&B\\B^*&A^t\end{array}\right),
\hskip .1cm Z=\left(\begin{array}{c} b \\ b^\dagger\end{array}\right),
\label{HbosZ}
 \end{eqnarray}
 \end{subequations}
where $A$ is an hermitian matrix, $B$ is symmetric and
$Z^\dagger=(b^\dagger,b)$, with $b$, $b^\dagger$ arrays of components $b_i$,
$b^\dagger_i$. The extended matrix $\mathcal{H}$ is hermitian and satisfies in
addition
 \begin{equation}
\bar{\mathcal{H}}\equiv\mathcal{T}\mathcal{H}^t\mathcal{T}=\mathcal{H}\,,\;\;\;
\mathcal{T}=\left(\begin{array}{cc}0&1\\1&0\end{array}\right)\,.\label{bas}
 \end{equation}
The boson commutation relations  $[b_i, b_j]=[b^\dagger_i, b^\dagger_j]=0$,
$[b_i,b^\dagger_j]=\delta_{ij}$, can be succinctly expressed as
 \begin{equation}
Z \, Z^\dagger-(Z^{\dagger\,t} \, Z^t)^t= \mathcal{M},\;\;\; \mathcal{M}= \left
(\begin{array}{cc} 1 & 0  \\  0 & -1
\end{array}\right)\,. \label{conmZ}
\end{equation}

It is well known that if the matrix $\mathcal{H}$ possesses only strictly {\it
positive} eigenvalues, the quadratic form (\ref{H}) can be diagonalized by
means of a standard linear Bogoliubov transformation for bosons preserving
Eqs.\ (\ref{conmZ}) \cite{RS.80}. This is the standard situation where
(\ref{H}) represents a stable system with a discrete positive spectrum, such as
a system of coupled harmonic oscillators. In general, however, and in contrast
with the fermionic case, it is not always possible to represent Eq.\ (\ref{H})
as a diagonal form in standard boson operators. The physical reason is obvious.
If ${\cal H}$ is not strictly positive, Eq.\ (\ref{H}) may represent the
Hamiltonian of systems like a free particle or a particle in a repulsive
quadratic potential ($H\propto p^2-q^2$) when expressed in terms of coordinates
and momenta, which do not possess a discrete spectrum. Nonetheless, one may
still attempt to write (\ref{H}) as a convenient diagonal form in suitable
operators, such that the ensuing equations of motion become decoupled and
trivial to solve.

Let us consider for this aim a general linear transformation \cite{RS.80,BR.86}
\begin{equation}
Z =\mathcal{W}\,Z'\,,\;\;\;\;
Z'=\left(\begin{array}{c}b'\\\bar{b'}\end{array}\right)\,,\label{ZZN}
\end{equation}
where $\bar{b}'_i$ is not necessarily the adjoint of $b'_i$, although $b'_i$,
$\bar{b}'_j$ are still assumed to satisfy the same boson commutation relations,
i.e., $Z'\bar{Z'}-(\bar{Z'}^t{Z'}^t)^t={\cal M}$, where
$\bar{Z'}\equiv(\bar{b}',b')=Z'^t\mathcal{T}$. Since
$Z^\dagger=\bar{Z}'\bar{\mathcal{W}}$, with
$\bar{\mathcal{W}}\equiv\mathcal{T}\mathcal{W}^t\mathcal{T}$, the matrix
$\mathcal{W}$ should then fulfill
\begin{equation}
\mathcal{W}{\cal M}\bar{\mathcal{W}}={\cal M}\,,\label{W}
\end{equation}
implying $\mathcal{W}^{-1}=\mathcal{M}\bar{\mathcal{W}}\mathcal{M}$. No
conjugation is involved in (\ref{W}). Note that $\bar{Z}\equiv
Z^t\mathcal{T}=Z^\dagger$ while in general $\bar{Z}'\neq
{Z'}^\dagger=\bar{Z}'\bar{\mathcal{W}}(\mathcal{W}^{\dagger})^{-1}$. If
$\bar{b}'={b'}^\dagger$ then $\bar{\mathcal{W}}=\mathcal{W}^\dagger$ (and
viceversa) and Eq.\ (\ref{ZZN}) reduces to a standard Bogoliubov transformation
for bosons \cite{RS.80,BR.86}. Eq.\ (\ref{ZZN}) allows to rewrite $H$ as
\begin{equation}
H=\half \bar{Z'}\mathcal{H}'Z',\;\;\;
\mathcal{H}'=\bar{\mathcal{W}}\mathcal{H}\mathcal{W}=
\left(\begin{array}{cc}A'&B'\\\bar{B}'&{A'}^t\end{array}\right)\,, \label{Hpw}
\end{equation}
where the relation (\ref{bas}) is preserved ($\bar{\mathcal{H}}'=\mathcal{H}'$,
implying $B',\bar{B}'$ symmetric), although in general
$\mathcal{H'}^\dagger\neq \mathcal{H'}$. Finding a representation where
$\mathcal{H'}$ is diagonal implies then an eigenvalue equation with ``metric''
$\mathcal{M}$, i.e.,
$\mathcal{H}\mathcal{W}=\mathcal{M}\mathcal{W}\mathcal{M}\mathcal{H'},$ which
can be recast as a standard eigenvalue equation for a {\it non-hermitian}
matrix $\tilde{\mathcal{H}}$:
\begin{equation}
\tilde{\mathcal{H}}\, \mathcal{W}= \mathcal{W}\,\tilde{\mathcal{H}}'\,,\;\;\;\;
\tilde{\mathcal{H}}\equiv\mathcal{M}\,\mathcal{H}=\left(\begin{array}{cc}A & B
\\ -B^* & -A^t\end{array}\right) \,. \label{Hd}
\end{equation}

This matrix is precisely that which determines the temporal evolution of the
system when $H$ is the Hamiltonian, as the Heisenberg equation of motion for
$b$, $b^\dagger$ is
\begin{equation}
i\frac{dZ}{dt}=-[H,Z]=\tilde{\mathcal{H}}Z\,.\label{eqm}
\end{equation}
Its solution for a time independent $\tilde{\mathcal{H}}$ is therefore
\begin{equation}
Z(t)=\mathcal{U}(t)Z(0)\,,\;\;\mathcal{U}(t)=\exp[-i\tilde{\mathcal{H}}t]\,,
 \label{Zt}\end{equation}
(or in general $\mathcal{U}(t)=T\exp[-i\int_0^t\tilde{\mathcal{H}}(t')dt']$,
where $T$ denotes time ordering). The eigenvalues of $\tilde{\mathcal{H}}$
characterize then the temporal evolution and can be {\it complex} in unstable
systems. Nevertheless, since
$\tilde{\mathcal{H}}^\dagger=\mathcal{H}\mathcal{M}=
\mathcal{M}\tilde{\mathcal{H}}\mathcal{M}$ and (Eq.\ (\ref{bas}))
\begin{equation}
\mathcal{T}\tilde{\mathcal{H}}^t\mathcal{T}=
-\mathcal{M}\tilde{\mathcal{H}}\mathcal{M}\,,\label{tmt}
\end{equation}
it is easily verified that the commutation relations (\ref{conmZ}) are always
preserved $\forall\,t\in\Re$, as
$\bar{\mathcal{U}}(t)\equiv\mathcal{T}\mathcal{U}^t\mathcal{T}=
\mathcal{U}^\dagger(t)$
and $\mathcal{U}(t)\mathcal{M}\bar{\mathcal{U}}(t)=\mathcal{M}$. Moreover, the
last identity remains valid also for {\it complex} times (although in this case
$\bar{\mathcal{U}}(t)\neq \mathcal{U}^\dagger(t)$), so that Eq.\ (\ref{Zt}) is
a particular example of the general transformation (\ref{ZZN}), becoming a
standard Bogoliubov transformation for bosons for $t\in\Re$.

Eq.\ (\ref{tmt}) implies that ${\rm Det}[\tilde{\mathcal{H}}^t-\lambda]={\rm
Det} [\tilde{\mathcal{H}}+\lambda]$, so that the eigenvalues of
$\tilde{\mathcal{H}}$ (the same as those of $\tilde{\mathcal{H}}^t$) always
come in pairs $(\lambda_i,\lambda_{\bar{i}})$ of {\it opposite} sign
($\lambda_{\bar{i}}=-\lambda_i)$. Eq.\ (\ref{tmt}) also entails that the
corresponding eigenvectors $W_i$ (columns of $\mathcal{W}$) satisfy the
orthogonality relations $\bar{W}_j\mathcal{M}W_i=-\bar{W}_i\mathcal{M}W_j=0$ if
$\lambda_i\neq -\lambda_j$, with $\bar{W}_i\equiv W_i^t\mathcal{T}$, which are
those required by Eq.\ (\ref{W}) (the required norm is
$\bar{W}_{\bar{i}}\mathcal{M}W_i=1$). In addition, for $\mathcal{H}$ hermitian,
${\rm Det}[\tilde{\mathcal{H}}-\lambda]^*= {\rm
Det}[\tilde{\mathcal{H}}^\dagger-\lambda^*]= {\rm
Det}[\tilde{\mathcal{H}}-\lambda^*]$, so that if $\lambda$ is an eigenvalue, so
is $\lambda^*$. Combined with (\ref{tmt}) this implies that if $W_i$ is
eigenvector with eigenvalue $\lambda_i$, $\mathcal{W}_{\bar{i}^*}\equiv
\mathcal{T}W_i^*$ is eigenvector with eigenvalue $-\lambda_i^*$. For
$\lambda_i$ {\it real}, the required norm can then be reduced to the usual one
for bosons \cite{RS.80}, $W_i^\dagger\mathcal{M}W_i=1$. However, for
$\lambda_i$ {\it complex}, the usual norm vanishes ($W_i^\dagger\mathcal{M}W_i=
\bar{W}_{\bar{i}^*}\mathcal{M}W_i=0$ as $\lambda_i\neq-\lambda_{\bar{i}^*}=
\lambda_i^*$) while the present one does not in general. Note finally that the
eigenvalues of $\tilde{\mathcal{H}}$ are the same as those of
$\tilde{\mathcal{H}}_s\equiv\sqrt{\mathcal{H}}\mathcal{M}\sqrt{\mathcal{H}}$.
When those of $\mathcal{H}$ are all {\it non-negative}, $\sqrt{\mathcal{H}}$
and hence $\tilde{\mathcal{H}}_s$ are {\it hermitian}, so that all eigenvalues
of $\tilde{\mathcal{H}}$ are {\it real}.

Let us assume now that the matrix $\tilde{\mathcal{H}}$ is {\it
diagonalizable}, such that a non-singular matrix $\mathcal{W}$ of eigenvectors
exists. Then $\bar{\mathcal{W}}\mathcal{M}\mathcal{W}$ will be non-singular,
and due to the orthogonality relations can be set equal to $\mathcal{M}$ if
eigenvectors are ordered and chosen such that
$\bar{W}_{\bar{j}}\mathcal{M}W_i=\delta_{ij}$. The ensuing $\mathcal{W}$ will
then satisfy Eqs.\ (\ref{W}) and (\ref{Hd}) with $\tilde{\mathcal{H}}'$
diagonal. Through the relation $\mathcal{H}'=\mathcal{M}\tilde{\mathcal{H}}'$
and Eq.\ (\ref{Hpw}) we obtain finally the diagonal representation
\begin{equation}
H=\sum_i\lambda_i(\bar{b}'_ib'_i+\half)\,,\label{Hdiag}
\end{equation}
where $b'_i=\bar{W}_{\bar{i}}\mathcal{M}Z$,
$\bar{b}'_i=Z^\dagger\mathcal{M}W_i$, with $W_i$, $W_{\bar{i}}$ the
eigenvectors with eigenvalues $\lambda_i$ and $-\lambda_i$ satisfying the
present norm ($\bar{W}_{\bar{i}}\mathcal{M}W_i=1$). If $\lambda_i$ is real, we
may choose $W_{\bar{i}}=\mathcal{T}W_i^*$ such that
$\bar{W}_{\bar{i}}=W_i^\dagger$ (with $W_i^\dagger \mathcal{M}W_i=1$) and hence
$\bar{b}'_i={b'_i}^\dagger$. Nonetheless, for complex $\lambda_i$,
$\bar{b}'_i\neq {b'_i}^\dagger$. Eq.\ (\ref{Hdiag}) remains, however,
physically meaningful, as the eigenvalues $\lambda_i$ determine the temporal
evolution. We immediately obtain from (\ref{Hdiag}) and (\ref{Zt}) the
decoupled evolution
 \begin{equation}
b'_i(t)=e^{-i\lambda_i t}b'_i(0),\;\;\bar{b}'_i(t)=
 e^{i\lambda_i t}\bar{b}'_i(0)\,,\label{Ztd}
\end{equation}
in all cases, together with the quadratic invariants $\bar{b}'_ib'_i=Z^\dagger
\mathcal{M}W_i\bar{W}_{\bar{i}}\mathcal{M}Z$. If all eigenvalues $\lambda_i$
are real and positive (with $\bar{b}'_i={b'_i}^\dagger$), we have the standard
case of a positive definite quadratic form. If all $\lambda_i$ are real but
some of them are {\it negative} (with $\bar{b}'_i={b'_i}^\dagger$), the system
is unstable in the sense that $H$ is no longer positive and does not possess a
minimum energy, but the spectrum is still discrete and the temporal evolution
(\ref{Zt}) remains stable. Finally, when some of the $\lambda_i$ are complex,
the evolution becomes unbounded, with $b'_i(t)$ ($\bar{b'_i}(t)$) increasing
(decreasing) exponentially for ${\rm Im}(\lambda_i)>0$ and increasing $t$.  In
these cases the sign of $\lambda_i$ in (\ref{Ztd}) depends on the choice of
operators and can be changed with the transformation $b'_i\rightarrow
-\bar{b}'_i$, $\bar{b}'_i\rightarrow b'_i$ (which preserves the commutation
relations) such that $\bar{b}'_ib'_i+\half \rightarrow -(\bar{b}'_ib'_i+\half)$
(for $\lambda_i$ real the sign can be fixed by the additional condition
$\bar{b}'_i={b'_i}^\dagger$). Cases where $\tilde{\mathcal{H}}$ is not
diagonalizable (which may arise when its eigenvalues are not all different) are
also dynamically unbounded as the temporal evolution determined by Eq.\
(\ref{Zt}) will contain terms proportional to some power of $t$ (times some
exponential; see example).

We may also express (\ref {H}) in terms of hermitian coordinates
$q=(b+b^\dagger)/\sqrt{2}$ and momenta $p=(b-b^\dagger)/(\sqrt{2}i)$,
satisfying $[p_i,p_j]=[q_i,q_j]=0$, $[q_i,p_j]=i\delta_{ij}$, as
\begin{subequations}\label{Hq}
\begin{eqnarray}
\!\!H&=&\half\sum_{i,j}T_{ij}p_ip_j+V_{ij}q_iq_j+U_{ij}\,q_i
p_j+U^t_{ij}\,p_iq_j\label{Hc}\\
&=&\half R^t \,\mathcal{H}_c \, R, \hskip .2cm \mathcal{H}_c= \left
(\begin{array}{cc}V&U\\U^t&T\end{array}\right), \hskip .1cm
R=\left(\begin{array}{c}q\\p\end{array}\right), \label{Hu}
\end{eqnarray}
\end{subequations}
where $V,T={\rm Re}(A\pm B)$ and $U={\rm Im}(B-A)$, with $T,V$ and
$\mathcal{H}_c$ {\it symmetric}. The corresponding transformation is
\begin{equation}
Z = \mathcal{S} \, R, \hskip 1.0cm
\mathcal{H}_c=\mathcal{S}^\dagger\mathcal{H}\mathcal{S}\,, \label{ZR}
\end{equation}
where $\mathcal{S}=\frac{1}{\sqrt{2}}(^{1\;\;\;i}_{1\;-i})$ is unitary and
satisfies $\mathcal{S}^\dagger=\mathcal{S}^t\mathcal{T}$. The commutation
relation for $R$ reads
\begin{equation}
RR^t-(RR^t)^t= \mathcal{M}_c,\;\;\;\mathcal{M}_c
=\mathcal{S}^\dagger\mathcal{M}\mathcal{S}=
\left(\begin{array}{cc}0&i\\-i&0\end{array}\right)
\,,\label{conmR}
\end{equation}
and the transformation (\ref{ZZN}) becomes
\begin{equation}
R=\mathcal{W}_cR'\,,\;\;\mathcal{W}_c\mathcal{M}_c\mathcal{W}_c^t=
\mathcal{M}_c\,,\label{Rw}
\end{equation}
where $\mathcal{W}_c=\mathcal{S}^\dagger\mathcal{W}\mathcal{S}$ and
$R'=(^{q'}_{p'})$ satisfies Eq.\ (\ref{conmR}). Note that $q',p'$ {\it will not
be hermitian} if $\mathcal{W}_c$ is complex. {\it Standard} linear canonical
transformations among hermitian coordinates and momenta correspond to
$\mathcal{W}_c$ {\it real}, which is equivalent to the condition
$\bar{\mathcal{W}}=\mathcal{W}^\dagger$ in (\ref{W}).

We may now rewrite (\ref{Hq}) as $H=\half {R'}^t\mathcal{H}'_cR'$, where
${\mathcal H}'_c=\mathcal{W}_c^t\mathcal{H}_c\mathcal{W}_c$ is symmetric
although not necessarily real. Finding a representation with $\mathcal{H}'_c$
{\it diagonal} implies then the non-standard eigenvalue problem
\begin{equation}
\tilde{\mathcal{H}}_c\mathcal{W}_c=\mathcal{W}_c\tilde{\mathcal{H}}'_c\,,\;\;
\tilde{\mathcal{H}}_c=\mathcal{M}_c\mathcal{H}_c=i\left(\begin{array}{cc}
     U^t&T \\-V & -U\end{array}\right)\,,\label{u}
\end{equation}
with $U'=0$ and $V',T'$ {\it diagonal} in $\tilde{\mathcal
H}'_c=\mathcal{M}_c\mathcal{H}'_c$, which leads to the coupled equations
$\tilde{\mathcal H}_cW_{ci}=-iV'_iW_{c\bar{i}}$, $\tilde{\mathcal
H}_cW_{c\bar{i}}=iT'_iW_{ci}$, for the columns of $\mathcal{W}_c$. The required
norm (Eq.\ (\ref{Rw})) is again $\bar{W}_{c\bar{i}}\mathcal{M}W_{ci}=1$. The
matrix $\tilde{\mathcal{H}}_c$ determines the evolution of $q,p$, as
$idR/dt=\tilde{\mathcal{H}}_cR$, and its eigenvalues are of course {\it the
same} as those of $\tilde{\mathcal{H}}$, as
$\tilde{\mathcal{H}}_c=\mathcal{S}^\dagger \tilde{\mathcal{H}}\mathcal{S}$. If
a matrix $\mathcal{W}_c$ (real or complex) satisfying (\ref{Rw})--(\ref{u})
exists, we obtain the diagonal form
\begin{equation}
H=\half\sum_i (T'_i{p'}^{\,2}_i+V'_i{q'}^{\,2}_i)\,,\;\;T'_iV'_i=\lambda_i^2\,,
\label{Hcdiagh}
\end{equation}
where $p'_i=-\bar{\mathcal{W}}_{ci}\mathcal{M}R$,
$q'_i=\bar{\mathcal{W}}_{c\bar{i}}\mathcal{M}R$ and $\lambda_i$ are the
eigenvalues of $\tilde{\mathcal{H}}$ or $\tilde{\mathcal{H}}_c$. For
$\lambda_i\neq 0$ we may always set $T'_i=V'_i=\lambda_i$ by a
scaling $p'_i\rightarrow s_i p'_i$, $q'_i\rightarrow q'_i/s_i$, where
$s_i=\sqrt[4]{V'_i/T'_i}$ can be complex, in which case we may choose
$W_{ci}=\mathcal{S}^\dagger(W_i+W_{\bar{i}})/\sqrt{2}$,
$W_{c\bar{i}}=i\mathcal{S}^\dagger(W_i-W_{\bar{i}})/\sqrt{2}$, with
$W_i,W_{\bar{i}}$ the eigenvectors of $\tilde{\mathcal{H}}$ with eigenvalues
$\pm\lambda_i$ satisfying $\bar{W}_{\bar{i}}\mathcal{M}\mathcal{W}_i=1$,
such that ${p'}_i^2+{q'}_i^2=2\bar{b}'_ib'_i+1$.
The ensuing operators $p'_i,q'_i$ will not be hermitian when $\lambda_i$ is
complex, but their evolution will still be given by the usual expressions
$q'_i(t)=q'_i(0)\cos(\lambda_i t)+p'_i(0)\sin(\lambda_it)$,
$p'_i(t)=p'_i(0)\cos(\lambda_i t)-q'_i(0)\sin(\lambda_it)$.

When $\tilde{\mathcal{H}}$ is {\it diagonalizable}, Eq.\ (\ref{Hcdiagh}) is
obviously equivalent to (\ref{Hdiag}) (with $Z'=\mathcal{S}R'$ for $T'=V'$).
However, Eq.\ (\ref{Hcdiagh}) is more general since it may also contain {\it
free particle terms} $\half T'_i {p'_i}^2$ when $\lambda_i=0$, which {\it
cannot} be written in the form (\ref{Hdiag}). In these cases the matrix
$\tilde{\mathcal{H}}$ {\it is not diagonalizable}, as easily recognized from
the ensuing linear evolution $p'_i(t)=p'(0)$, $q'_i(t)=q'_i(0)+tT'_ip'_i(0)$,
having a degenerate eigenvalue 0.  Nonetheless, it should be emphasized that
{\it it is not always possible} to represent Eq.\ (\ref{Hq}) in the diagonal
form (\ref{Hcdiagh}), as non-diagonalizable cases where {\it no} eigenvalue of
$\tilde{\mathcal{H}}$ vanishes, {\it also exist} (see example). Let us also
remark that if one considers just {\it hermitian} $q'_i,p'_i$ in
(\ref{Hcdiagh}),  with $T'_i,V'_i$  {\it real}, the eigenvalues $\lambda_i$ of
$\tilde{\mathcal{H}}$ are either real ($T'_iV'_i\geq 0$) or purely imaginary
($T'_iV'_i<0$). Thus, quadratic forms whose matrix $\tilde{\mathcal{H}}$
possesses full complex eigenvalues (see example) {\it cannot be written in the
diagonal form (\ref{Hcdiagh}) unless non-hermitian coordinates and momenta
$q',p'$ are admitted}.

The following example clearly illustrates the previous situations. Let us
consider the Hamiltonian
\begin{subequations}
\label{heg}
\begin{eqnarray}
H&=&\sum_{\nu=\pm}\varepsilon_\nu(b^\dagger_\nu b_\nu+\half)
+\Delta(b_+ b_-+b^\dagger_+ b^\dagger_-) \label{hej}\\
&=&\half\sum_{\nu=\pm}\varepsilon_\nu(p_\nu^2+q_\nu^2)
+\Delta(q_+q_--p_+p_-)\,,\label{hejq}
\end{eqnarray}
\end{subequations}
which represents two boson modes interacting through a BCS-like pairing term.
We assume $\varepsilon_+>\varepsilon_->0$, and write
$\varepsilon_\pm=\varepsilon\pm\gamma$, with $\varepsilon>0$,
$0<\gamma<\varepsilon$. The eigenvalues of the ensuing matrix $\mathcal{H}$ (or
$\mathcal{H}_c$), two-fold degenerate, are
\begin{equation}
\sigma_{\pm}=\varepsilon\pm\sqrt{\gamma^2+\Delta^2}\,,
\end{equation}
which are both positive only for
$|\Delta|<\sqrt{\varepsilon^2-\gamma^2}=\sqrt{\varepsilon_+\varepsilon_-}$ (the
condition for a positive mass and potential tensor in (\ref{hejq})). However,
the four eigenvalues of $\tilde{\mathcal{H}}= \mathcal{M}\mathcal{H}$ are
\begin{equation}
\lambda^{\pm}_\nu=
\pm[\nu\gamma+\sqrt{\varepsilon^2-\Delta^2}],\;\;\nu=\pm\,,\label{lambda}
\end{equation}
which are real for $|\Delta|\leq\varepsilon=(\varepsilon_++\varepsilon_-)/2$.
Thus, if $\sqrt{\varepsilon^2-\gamma^2}<|\Delta|<\varepsilon$, $H$ is no longer
positive definite ($\sigma_-<0$), but all eigenvalues $\lambda^{\pm}_\nu$
remain {\it real} (and distinct) implying that the temporal evolution is still
{\it bounded} (quasiperiodic). However, for $|\Delta|>\varepsilon$, all
eigenvalues are complex (with {\it non-zero} real part if $\gamma\neq 0$) and
the evolution becomes unbounded.

Let us obtain now the diagonal representation of $H$. It is sufficient to
consider in (\ref{W}) a BCS-like transformation for bosons of the form
\begin{equation}
b_\nu=ub'_\nu-v\bar{b}'_{-\nu},\;\;b^\dagger_\nu=u\bar{b}'_\nu-vb'_{-\nu}\,,
\label{bprime}
\end{equation}
which corresponds to $q_\nu=uq'_\nu-vq'_{-\nu}$, $p_\nu=up'_\nu+vp'_{-\nu}$.
The commutation relations are preserved if $u^2-v^2=1$
($\mathcal{W}\mathcal{M}\bar{\mathcal{W}}=\mathcal{M}$) and the inverse
transformation ($\mathcal{M}\bar{\mathcal{W}}\mathcal{M}$) is obtained for
$v\rightarrow -v$ ($b'_\nu=ub_\nu+vb^\dagger_{-\nu}$,
$\bar{b}'_\nu=ub^\dagger_\nu+vb_{-\nu}$). Now, for
\begin{equation}
\left(\begin{array}{c}u\\v\end{array}\right)=
\sqrt{\frac{\varepsilon\pm\alpha}{2\alpha}}\,, \;\;\;
\alpha=\sqrt{\varepsilon^2-\Delta^2}\,,
\end{equation}
where we assume $\alpha\neq 0$ ($|\Delta|\neq\varepsilon$) and signs in square
roots are to be chosen such that $2\alpha uv=\Delta$, we may express $H$ as a
sum of two independent modes,
\begin{equation}
H=\sum_{\nu=\pm}\lambda_\nu(\bar{b}'_\nu b'_\nu+\half)=
\half\sum_{\nu=\pm}\lambda_{\nu}({p'_\nu}^2+{q'_\nu}^2)\,,\label{go}
 \end{equation}
where $\lambda_\nu\equiv\lambda_\nu^+$. If $|\Delta|<\varepsilon$, $u,v$ are
both real, so that $\bar{b}'_\nu={b'}_\nu^{\dagger}$, with $q'_\nu,p'_\nu$,
hermitian, while if $|\Delta|>\varepsilon$, $u,v$ are {\it complex}, implying
$\bar{b}'_i\neq {b'}_i^\dagger$ and $q'_i$, $p'_i$ no longer hermitian.
Instead, $(\lambda_{\nu})^*=-\lambda_{-\nu}$ and $u^*=iv$ (with ${\rm
Im}(\alpha)>0$ for $\Delta>0$), entailing ${b_\nu'}^{\dagger}=ib'_{-\nu}$, 
${{\bar{b}}'_\nu}{}^{\dagger}=i\bar{b}'_{-\nu}$ and
${q'_\nu}^{\dagger}=iq'_{-\nu}$, ${p'}_{\nu}^{\dagger}=-ip'_{-\nu}$. Note that
in this case the usual norm vanishes ($|u|^2-|v|^2=0$) but the present one
remains unchanged ($u^2-v^2=1$ still holds).

If $|\Delta|<\sqrt{\varepsilon^2-\gamma^2}$, $\lambda_{\pm}>0$, so that both
modes have a discrete positive spectrum. However, if
$\sqrt{\varepsilon^2-\gamma^2}<|\Delta|<\varepsilon$, $\lambda_+>0$ but
$\lambda_-<0$, so that the spectrum of the lowest mode, though still discrete,
becomes {\it negative}, implying that $H$ has no longer a minimum energy.
Careful should be taken here to select the correct eigenvalue in
(\ref{lambda}), as $\tilde{\mathcal{H}}$ still has two positive eigenvalues
($\lambda^-_{-}>0$). Note also that for
$|\Delta|=\sqrt{\varepsilon^2-\gamma^2}$, $\lambda_-^{\pm}=0$, reflecting the
onset of the instability, but $\tilde{\mathcal{H}}$ is still {\it
diagonalizable}, as $u,v$ remain finite. The lowest mode in (\ref{go}) has here
a single degenerate eigenvalue 0. Finally, for $|\Delta|>\varepsilon$, the
operators $b'_\nu$, $\bar{b}'_\nu$ represent complex modes with an
exponentially increasing or decreasing evolution. The evolution of the original
operators $b_\nu,b^\dagger_\nu$ for any $|\Delta|\neq\varepsilon$ can be
immediately obtained from (\ref{Ztd}) and (\ref{bprime}) and is given by
\begin{equation}
b_\nu(t)=e^{-i\lambda_\nu t}[b_\nu+v(1-e^{2i\alpha t})
(vb_\nu+ub^\dagger_{-\nu})]\,,\label{bt}
\end{equation}
where $b_\nu\equiv b_\nu(0)$, $b_\nu^\dagger\equiv b_\nu^\dagger(0)$, with
$b_\nu^\dagger(t)=[b_\nu(t)]^\dagger$. It becomes clearly unbounded for
$|\Delta|>\varepsilon$.

For $|\Delta|=\varepsilon$, $\tilde{\mathcal{H}}$ {\it is not diagonalizable},
even though its eigenvalues $\lambda_\nu^{\pm}$ are in this case {\it all real
and non-zero} (but degenerate), and $H$ cannot be written in the form
(\ref{go}). However, the time evolution can still be obtained from (\ref{bt})
taking the limit $\alpha\rightarrow 0$, which leads to
\begin{equation}\label{bs}
b_\nu(t)=e^{-i\nu \gamma t}[(1-it\varepsilon)b_\nu-it\Delta
b_{-\nu}^\dagger]\,.
\end{equation}
The factor $t$ confirms that the evolution equations cannot be fully decoupled
in this case, while the exponential multiplying this factor shows that they do
not arise from a free particle term either. We may, however, rewrite $H$ in
this case (assuming for instance $\Delta=\varepsilon$) as
 \begin{equation}\label{hdec}
 H=\gamma(\bar{b}^s_+b^s_+-\bar{b}^s_-b^s_-)
 +2\Delta\bar{b}^s_-\bar{b}^s_+\,,
\end{equation}
where $b_\nu=(b^s_\nu+\bar{b}^s_{-\nu})/\sqrt{2}$,
$b_{\nu}^\dagger=(\bar{b}^s_{\nu}-b^s_{-\nu})/\sqrt{2}$, with
${b^s_\nu}{}^\dagger=-b^s_{-\nu}$,
${\bar{b}^s_{\nu}}{}^\dagger=\bar{b}^s_{-\nu}$, also satisfy boson commutation
relations. In the form (\ref{hdec}) $H$ is ``maximally decoupled'', in the
sense that the evolution equations for $\bar{b}^s_\nu$ are fully {\it
decoupled}, while those of $b^s_\nu$ are coupled just to $\bar{b}^s_{-\nu}$.
This leads to $\bar{b}^s_\nu(t)=e^{i\nu\gamma t}\bar{b}^s_\nu$, $b^s_\nu(t)=
e^{-i\nu\gamma t}(b^s_\nu- 2it\Delta\bar{b}^s_{-\nu})$. Eq.\ (\ref{bs}) can
also be obtained from these expressions. The associated invariants in this case
are $\bar{b}^s_-\bar{b}^s_+$ and $\bar{b}^s_+b^s_+-\bar{b}^s_-b^s_-$, i.e., the
two terms in (\ref{hdec}), which are mutually commuting.

If $b_\nu,b^\dagger_\nu$ were fermion operators, Eq.\ (\ref{hej}) would
represent essentially a generic term of the standard BCS approximation to a
pairing Hamiltonian \cite{RS.80} [$H_{\scriptscriptstyle BCS}=
\sum_{k,\nu}\varepsilon_{k\nu}b^\dagger_{k\nu}b_{k\nu}+\sum_k \Delta_k(b_{k+}
b_{k-}+b^\dagger_{k-}b^\dagger_{k+})$, where $k\pm$ denote time reversed
states, $\Delta_k$ the BCS gap, $b_{k\nu},b^\dagger_{k\nu}$ fermion operators
and the splitting between $\varepsilon_{k\pm}$ may represent the effect of a
Zeeman coupling to a magnetic field]. In the fermionic case, Eq.\ (\ref{hej})
(with $\half\rightarrow-\half$) can be written as $\sum_\nu\lambda_\nu
({b'}^\dagger_{\nu}b'_\nu-\half)$ $\forall$ $\Delta$, where
$\lambda_\nu=\nu\gamma+\alpha$, with $\alpha=\sqrt{\varepsilon^2+\Delta^2}$,
are the quasiparticle energies and $b'_\nu, {b'}^\dagger_\nu$ quasiparticle
fermion operators defined by $b_\nu=ub'_\nu+\nu v{b'}^\dagger_{-\nu}$, with
$u,v=\sqrt{(\alpha\pm\varepsilon)/2\alpha}$. The analogous boson problem is, in
contrast, stable just for limited values of $\Delta$, as the latter {\it
decreases} (rather than increases) the ``quasiparticle energies''
$\lambda_\nu$. The onset of complex frequencies occurs finally when
$\lambda_-=-\lambda_+$.

Let us also mention that in general, when $\mathcal{H}$ is not positive regions
of dynamical stability may also arise between fully unstable regions. For
instance, if a perturbation $\kappa(b^\dagger_+b_-+b^\dagger_{-}b_+)$ is added
to (\ref{heg}), the eigenvalues of $\mathcal{H}$ and $\tilde{\mathcal{H}}$
become $\sigma^\pm_\nu=\varepsilon+\nu\sqrt{\gamma^2+(\Delta\pm\kappa)^2}$ and
$\lambda_{\nu}^\pm=\pm\sqrt{\tilde{\lambda}_\nu^2-\kappa^2
(\varepsilon^2/\gamma^2-1)}$, with
$\tilde{\lambda}_\nu=\nu\gamma+\sqrt{\Delta_c^2-\Delta^2}$ and $\Delta_c=
\varepsilon^2(1+\kappa^2/\gamma^2)$. Those of $\mathcal{H}$ are split, and
assuming $\kappa$ small such that $\mathcal{H}$ is positive at $\Delta=0$, the
two lowest ones $\sigma_{-}^{\pm}$ become negative at different values
$\Delta_{c\pm}=\sqrt{\varepsilon^2-\gamma^2}\pm|\kappa|$. In such a case
$\lambda_{-}^\pm$ becomes {\it imaginary} for
$\Delta_{c-}<|\Delta|<\Delta_{c+}$, but returns again to {\it real values} for
$\Delta_{c+}<|\Delta|<\Delta_{c}$ if
$|\kappa|<\gamma^2/\sqrt{\varepsilon^2-\gamma^2}$, exhibiting a reentry of
dynamical stability. Finally, both $\lambda_{\pm}$ become full complex for
$|\Delta|>\Delta_{c}$. A diagonal representation of the general form (\ref{go})
is feasible except at the critical values $\Delta_{c\pm}$ and $\Delta_{c}$.

In summary, we have extended the standard methodology employed for
diagonalizing an hermitian quadratic bosonic form, employing generalized
quasiparticle boson-like operators for describing unstable cases with arbitrary
complex frequencies. In this way the operators exhibiting an exponentially
increasing or decreasing temporal evolution are explicitly identified, together
with the associated quadratic invariants, allowing for a precise
characterization of the system evolution in the presence of general
instabilities. While positive definite forms can be considered completely
stable, those which are not positive but whose matrix $\tilde{\mathcal{H}}$ is
diagonalizable and has only {\it real} eigenvalues, can still be considered
{\it dynamically} stable, as the temporal evolution remains quasiperiodic, in
contrast with the case where $\tilde{\mathcal{H}}$ has complex eigenvalues or
is non-diagonalizable. Finally, we have seen that a BCS-like hamiltonian for
bosons can be completely stable, just dynamically stable, or unstable depending
on the values of the gap parameter, and requires the present generalized
approach for a diagonal representation valid for large gaps. Moreover, it also
shows that cases where $\tilde{\mathcal{H}}$ is non-diagonalizable are not
necessarily associated with zero frequencies or free particle terms, and may
arise even if all its eigenvalues are non-zero. For such cases the evolution
equations cannot be fully decoupled.

RR and AMK are supported by CIC of Argentina.

\end{document}